\documentclass[conference, a4paper]{IEEEtran}

\usepackage{cite}
\usepackage{color}
\usepackage{threeparttable}
\usepackage{graphicx}
\usepackage{picinpar}
\usepackage[cmex10]{amsmath}
\usepackage{amsmath,amsfonts,amssymb}
\usepackage{subfigure}
\usepackage{changepage}
\usepackage{algorithm}
\usepackage{caption}
\usepackage{algpseudocode}
\usepackage{stfloats}
\usepackage{bm}
\usepackage{enumerate}

\begin{document}
\newtheorem{lemma}{Lemma}
\newtheorem{corol}{Corollary}
\newtheorem{theorem}{Theorem}
\newtheorem{proposition}{Proposition}
\newtheorem{definition}{Definition}
\newcommand{\e}{\begin{equation}}
\newcommand{\ee}{\end{equation}}
\newcommand{\eqn}{\begin{eqnarray}}
\newcommand{\eeqn}{\end{eqnarray}}
%  调整公式间距
\newenvironment{shrinkeq}[1]
{ \bgroup
\addtolength\abovedisplayshortskip{#1}
\addtolength\abovedisplayskip{#1}
\addtolength\belowdisplayshortskip{#1}
\addtolength\belowdisplayskip{#1}}
{\egroup\ignorespacesafterend}
\title{Broadband Channel Estimation for Intelligent \\ Reflecting Surface Aided mmWave \\ Massive MIMO Systems}

\author{\IEEEauthorblockN{Ziwei~Wan\IEEEauthorrefmark{1}\IEEEauthorrefmark{2}, Zhen~Gao\IEEEauthorrefmark{2}\IEEEauthorrefmark{1} and Mohamed-Slim Alouini\IEEEauthorrefmark{3}}\\
\vspace{-2mm}
\IEEEauthorblockA{\IEEEauthorrefmark{1}School of Information and Electronics, Beijing Institute of Technology, Beijing 100081, P. R. China}\\
\vspace{-4mm}
\IEEEauthorblockA{\IEEEauthorrefmark{2}Advanced Research Institute of Multidisciplinary Science, Beijing Institute of Technology, Beijing 100081, P. R. China}\\
\vspace{-4mm}
\IEEEauthorblockA{\IEEEauthorrefmark{3}Electrical Engineering Program, Division of Physical Sciences and Engineering,\\ King Abdullah University of Science and Technology (KAUST), Thuwal, Makkah Province, Saudi Arabia}
\vspace{-4mm}\\
Email: ziweiwan@bit.edu.cn, gaozhen16@bit.edu.cn, slim.alouini@kaust.edu.sa}

%\author{\IEEEauthorblockN{Ziwei~Wan\IEEEauthorrefmark{1}\IEEEauthorrefmark{2} and Zhen~Gao\IEEEauthorrefmark{1}\IEEEauthorrefmark{2}}\\
%\vspace{-2mm}
%\IEEEauthorblockA{\IEEEauthorrefmark{1}School of Information and Electronics, Beijing Institute of Technology, Beijing 100081, P. R. China}\\
%\vspace{-4mm}
%\IEEEauthorblockA{\IEEEauthorrefmark{2}Advanced Research Institute of Multidisciplinary Science, Beijing Institute of Technology, Beijing 100081, P. R. China}
%\vspace{-4mm}\\
%Email: ziweiwan@bit.edu.cn, gaozhen16@bit.edu.cn}

\maketitle

\begin{abstract}
This paper investigates the broadband channel estimation (CE) for intelligent reflecting surface (IRS)-aided millimeter-wave (mmWave) massive MIMO systems.
The CE for such systems is a challenging task due to the large dimension of both the active massive MIMO at the base station (BS) and passive IRS.
To address this problem, this paper proposes a compressive sensing (CS)-based CE solution for IRS-aided mmWave massive MIMO systems,
whereby the angular channel sparsity of large-scale array at mmWave is exploited for improved CE with reduced pilot overhead.
Specifically, we first propose a downlink pilot transmission framework.
By designing the pilot signals based on the prior knowledge that the line-of-sight dominated BS-to-IRS channel is known, the high-dimensional channels for BS-to-user and IRS-to-user can be jointly estimated based on CS theory.
Moreover, to efficiently estimate broadband channels, a distributed orthogonal matching pursuit algorithm is exploited, where the common sparsity shared by the channels at different subcarriers is utilized.
Additionally, the redundant dictionary to combat the power leakage is also designed for the enhanced CE performance.
Simulation results demonstrate the effectiveness of the proposed scheme.
\end{abstract}


\IEEEpeerreviewmaketitle

\section{Introduction}
Inspired by the new paradigm of smart and reconfigurable wireless environment \cite{Europe}, the intelligent reflecting surface (IRS) has been considered as a key enabling technology for 5G beyond and even 6G \cite{Europe, Fellow3, AA, LSCE, MSlim, RZhang2, XYuan}. On the one hand, with a number of reconfigurable elements built by advanced materials \cite{Fellow3}, IRS can reflect the electromagnetic signals in a desired mode based on the channel state information (CSI) to achieve better communication capacity \cite{RZhang2, MSlim, LSCE}. On the other hand, distinctive from the existing similar technologies like relays or multiple-input multiple-output (MIMO) beamforming, the passive IRS needs neither energy-hungry active radio frequency (RF) devices nor complicated baseband (BB) signal processing modules, which meets the demand for the green and energy-efficient technologies in the future communication systems.

IRS cooperating with the emerging MIMO technology has attracted extensive attention \cite{RZhang2, MSlim}.
In \cite{RZhang2}, by maximizing the receive signal-to-interference-plus-noise ratio (SINR), the reflection-coefficients at the IRS are designed based on the convex optimization methods.
A similar scenario is also considered in \cite{MSlim}, where the phases of the IRS that maximize the SINR are obtained by using projected gradient descent.
%Further, it has been pointed out in \cite{MSlim} that it would be harmful to serve multiple users by a single IRS if there only exists a line-of-sight (LoS) channel between the BS and IRS.

The solutions and analysis in most previous work like \cite{RZhang2, MSlim} are based on the perfect knowledge of all the channels of interest. Nevertheless, the CSI acquisition is a non-trivial task in IRS-aided massive MIMO systems due to the high-dimensional channels and passive property of IRS \cite{Fellow3, AA, LSCE, XYuan}. The channel estimation (CE) for IRS-aided MIMO systems have been investigated recently.
In \cite{LSCE}, a CE paradigm based on the conventional least square (LS) estimator and a low-complexity beamforming design have been proposed for IRS assisted MISO systems. The estimation of the cascaded BS-IRS-user channel is elaborated in \cite{XYuan}, where a sophisticated algorithm based on sparse matrix factorization and matrix completion has been introduced to solve the CE problem. To obtain the reliable beamforming design in the device-to-device communication via IRS, the authors of \cite{AA} have proposed a new architecture of IRS, where a few RF-chains are equipped at the IRS to sound the channel in real time. The compressive sensing as well as deep learning techniques are adopted for the beamforming design.
However, the solutions in \cite{AA} require the IRS to deploy RF-chains, which contradicts the intention to introduce the passive IRS for energy-efficient communications.
Moreover, the CE solution in \cite{LSCE} suffers from prohibitive pilot overhead because of the inherent limitation of the conventional CE technique.
Besides, the frequency selective fading channel property for broadband systems \cite{GaoCL, GaoTSP} is seldom considered in the existing work on the CE for IRS-aided systems.

In this paper, we propose a compressive sensing (CS)-based broadband CE solution for IRS-aided mmWave massive MIMO systems (see Fig. \ref{fig:model}), where the orthogonal frequency division multiplex (OFDM) is adopted to combat the frequency selective fading property of the channels. The hybrid analog-digital MIMO architecture \cite{GaoCL} at the BS and the RF-chain-free IRS are considered to reduce the power consumption.
We first propose a downlink pilot transmission framework to sound the channels.
Then, by designing the pilot signals based on the prior knowledge that the line-of-sight (LoS) dominated BS-to-IRS channel is known, we can jointly estimate the BS-to-user and IRS-to-user channels based on CS theory with reduced training overhead.
A distributed orthogonal matching pursuit algorithm which utilizes the common sparsity shared by the channels at different subcarriers is presented to efficiently estimate the channels.
Moreover, we design the redundant dictionary to combat the power leakage for the enhanced CE performance. Numerical results verify the effectiveness of the proposed CE scheme.
{\it Our work is an initial attempt to discuss the broadband channel estimation for IRS-aided mmWave massive MIMO with the hybrid architecture.}

\textit{Notations}:
Column vectors and matrices are denoted by lower- and upper-case boldface letters, respectively.
${\left(  \cdot  \right)\!^T}$, ${\left( \cdot \right)\!^H}$ and ${\left(  \cdot  \right)\!^{\dag}}$ denote the transpose, conjugate transpose and the pseudo-inverse, respectively.
The imaginary unit is defined as $j = \sqrt{-1}$.
$\mathbb{C}^{M \times N}$ is the set of $M \times N$ matrices with complex-valued entries.
${[ \cdot ]_{i}}$ and ${[ \cdot ]_{i,j}}$ represent the ${i}$-th element of a vector and ${i}$-th row, ${j}$-th column element of a matrix, respectively.
$[\bf A]_{\cal I}$ denotes the submatrix consisting of the columns of $\bf{A}$ indexed by the set $\cal I$.
${\bf 1} \in \mathbb{C}^{M \times N}$ is an all-one matrix of size $M \times N$.
${\left\| \cdot \right\|_p}$, ${\rm{diag(}} \cdot {\rm{)}}$ and $ \otimes $ represent the $l_p$-norm of a vector, diagonalization and Kronecker product, respectively.
${\rm{supp(}} \cdot {\rm{)}}$ is the support set, which is constructed by the indices of the non-zero elements of a sparse vector.

\section{System Model}
\begin{figure}[t]
     \centering
     \includegraphics[width=6cm, keepaspectratio]%
     {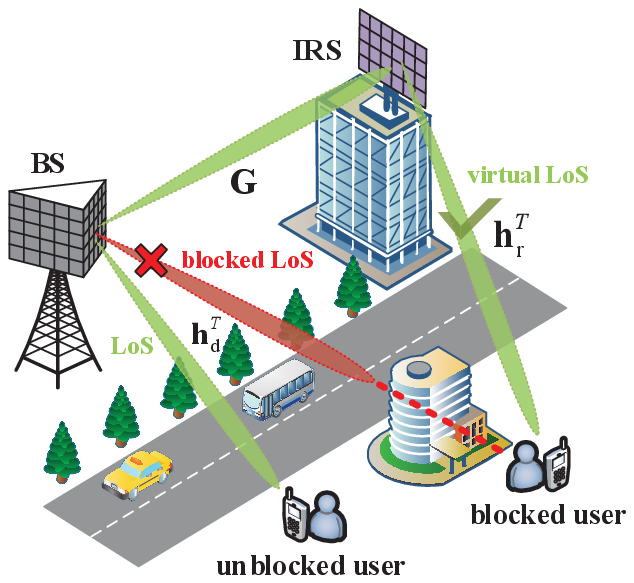}
     \caption{IRS-aided mmWave massive MIMO systems.}
     \label{fig:model}
%     \vspace*{-4.5mm}
\end{figure}

Consider a broadband mmWave cellular system consisting of one BS and one IRS, as shown in Fig. \ref{fig:model}. It can be seen in Fig. \ref{fig:model} that IRS can provide a {\it virtual LoS} path for the users when the practical LoS path is blocked by the possible obstacles or human bodies. This behaviour can help to facilitate the mmWave transmissions which are limited by the high penetration loss.
We consider that both BS and IRS are equipped with half-wavelength-spaced uniform planar arrays (UPAs), where the numbers of antennas are $M = M_x \times M_y$ and $N = N_x \times N_y$, respectively. Single-antenna users are served within the cell.
We consider the widely-used hybrid architecture of mmWave massive MIMO at the BS, i.e., only $N_{\rm RF} \ll M$ RF chains are equipped at the BS and each of them is connected to $M$ antennas through $M$ phase shifters.
To combat the frequency-selective fading property of the channels, OFDM system with $K$ subcarriers and sampling space $T_s$ is adopted in the system.
We focus on the downlink CE throughout the paper. The delay-domain channels from the BS to the user, from the BS to the IRS, and from the IRS to the user are denoted by ${\bf h}_{{\rm{d}}}^T (\tau)$, ${{\bf{G}}} (\tau)$ and ${\bf{h}}_{{\rm{r}}}^T (\tau)$, respectively. Taking ${{\bf{G}}} (\tau)$ as an example, we can model it as
\begin{align}
\label{equ:G}
{{\bf{G}}} & (\tau ) = \frac{{{\alpha _{{\rm{g,0}}}}}}{{{\rho _{{\rm{g,0}}}}}}{{\bf{a}}_N}(\theta _{{\rm{g}},0}^{\rm{r}},\varphi _{{\rm{g}},0}^{\rm{r}}){\bf{a}}_M^H(\theta _{{\rm{g}},0}^{\rm{t}},\varphi _{{\rm{g}},0}^{\rm{t}}) \times {p}(\tau - \tau _{{\rm{g}},0} ) \nonumber \\
& + \sum\limits_{l = 1}^{{L_{\rm{g}}}} {\frac{{{\alpha _{{\rm{g,}}l}}}}{{{\rho _{{\rm{g,}}l}}}}} {{\bf{a}}_N}(\theta _{{\rm{g}},l}^{\rm{r}},\varphi _{{\rm{g}},l}^{\rm{r}}){\bf{a}}_M^H(\theta _{{\rm{g}},l}^{\rm{t}},\varphi _{{\rm{g}},l}^{\rm{t}}) \times {p}(\tau  - {\tau _{{\rm{g}},l}}) \text{,}
\end{align}
where $L_{\rm g}$ is the number of NLoS paths, $\theta _{{\rm{g}},l}^{\rm{t}}$ ($\theta _{{\rm{g}},l}^{\rm{r}}$) and $\varphi _{{\rm{g}},l}^{\rm{t}}$ ($\varphi _{{\rm{g}},l}^{\rm{r}}$) are the horizontal and vertical component of the angle of departure (arrival, AoD/AoA) of the $l$-th path, respectively, ${{\alpha _{{{\rm{g}},l}}}}$ and ${{\rho _{{{\rm{g}},l}}}}$ represent the complex gain and the large-scale fading coefficient corresponding to the $l$-th path, respectively, $\tau_{{\rm{g}},l}$ is the delay offset of the $l$-th path, and $p(\tau)$ is the pulse shaping filter function with the sampling space $T_s$. Note that $l=0$ for LoS path and $l>0$ for $l$-th NLoS path. The steering vector of the half-wavelength-spaced UPA ${\bf{a}}_M(\theta,\varphi) \in {\mathbb{C}^{M \times 1}}$ is defined by
\begin{align}
\label{equ:a_UPA}
{{\bf{a}}_M}(\theta ,\varphi ) = & \frac{1}{{\sqrt M }} {[1,{e^{j\pi \sin \theta }},...,{e^{j\pi ({M_x} - 1)\sin \theta }}]^T} \nonumber \\
& \otimes {[1,{e^{j\pi \cos \theta \sin \varphi }},...,{e^{j\pi ({M_y} - 1)\cos \theta \sin \varphi }}]^T} \text{,}
\end{align}
and ${\bf{a}}_N(\theta,\varphi) \in {\mathbb{C}^{N \times 1}}$ can be similarly derivated.

In OFDM system, the delay-domain channel $\bf{G}(\tau)$ can be transformed to $K$ frequency-domain subchannels ${\bf{G}}_{k}$, $k=1,...,K$ via discrete Fourier transformation (DFT) and each subchannel can be treated as a flat-fading channel, that is
\begin{align}
{{\bf{G}}_{k}} & = \sum\limits_{d = 0}^{{N_{{\rm{CP}}}} - 1} {{\bf{G}}(d{T_s})} {e^{j\frac{{2\pi (k-1)}}{K}d}} \nonumber \\
& = \underbrace {{{g _{{\rm{g}},0,k}}}{{\bf{a}}_N}(\theta _{{\rm{g}},0}^{\rm{r}},\varphi _{{\rm{g}},0}^{\rm{r}}){\bf{a}}_M^H(\theta _{{\rm{g}},0}^{\rm{t}},\varphi _{{\rm{g}},0}^{\rm{t}})}_{{{\bf{G}}_{{\rm{L}},k}}} \nonumber \\
&\quad + \underbrace {\sum\limits_{l = 1}^{{L_{\rm{g}}}} {{g _{{\rm{g}},l,k}}} {{\bf{a}}_N}(\theta _{{\rm{g}},l}^{\rm{r}},\varphi _{{\rm{g}},l}^{\rm{r}}){\bf{a}}_M^H(\theta _{{\rm{g}},l}^{\rm{t}},\varphi _{{\rm{g}},l}^{\rm{t}})}_{{\bf{G}}_{{\rm{N}},k}} \text{,}
\label{equ:Gf1}
\end{align}
where $N_{\rm CP}$ is the length of cyclic prefix (CP) in OFDM system to avoid the inter-symbol interference, and ${g _{{\rm{g}},l,k}} = \frac{{{\alpha _{{\rm{g,}}l}}}}{{{\rho _{{\rm{g,}}l}}}}\sum\nolimits_{d = 0}^{{N_{{\rm{CP}}}} - 1} {{p}(d{T_s} - {\tau _{{\rm{g}},l}}){e^{j\frac{{2\pi (k-1)}}{K}d}}} $. Note that in (\ref{equ:Gf1}), we decompose the frequency-domain subchannel into two parts: the LoS part ${{{\bf{G}}_{{\rm{L}},k}}} \in {\mathbb{C}^{N \times M}}$ and the non-LoS (NLoS) part ${{\bf{G}}_{{\rm{N}},k}} \in {\mathbb{C}^{N \times M}}$. Thus, the frequency-domain channels can be compactly expressed as

\begin{equation}
\label{equ:Gf2}
{{\bf{G}}_{k}} = {{\bf{G}}_{{\rm{L}},k}} + {{\bf{G}}_{{\rm{N}},k}} \text{.}
\end{equation}

Follow the similar procedure from (\ref{equ:G}) to (\ref{equ:Gf1}), the frequency-domain channels associated with ${\bf h}_{{\rm{d}}}^T (\tau)$ and ${\bf{h}}_{{\rm{r}}}^T (\tau)$ can also be calculated as ${\bf{h}}_{{\rm{d}},k}^T \in \mathbb{C}^{1 \times M}$ and ${\bf{h}}_{{\rm{r}},k}^T \in \mathbb{C}^{1 \times N}$, $k = 1,...,K$, respectively.

\section{Proposed Channel Estimation Technique}
As we mentioned above, unlike the IRS with receive RF chains in \cite{AA}, the IRS applied in our system do not equip any receive RF chain, so it is almost impossible to estimate the BS-to-IRS channel ${{\bf{G}}_{k}}$ via the conventional pilot-aided CE schemes. To begin with, we first consider the estimation of ${\bf{h}}_{{\rm{d}},k}^T$ and ${\bf{h}}_{{\rm{r}},k}^T$ under the assumption that ${{\bf{G}}_{k}}$ is known, and then we will justify the reasonability of this assumption.

\subsection{Downlink Pilot Transmission}
To conduct downlink CE, BS will broadcast the pilot signals which are identical for all users to sound the channels. Taking into account the hybrid architecture at the BS, The transmit pilot signal can be written as
\begin{equation}
\label{equ:S}
{{\bf{s}}_{i,k}} = {{\bf{F}}_{{\rm{RF}},i}}{{\bf{f}}_{{\rm{BB}},i,k}} \text{,}
\end{equation}
where ${\bf{s}}_{i,k} \in \mathbb{C}^{M \times 1}$ is the transmit pilot signal dedicated to the $k$-th subcarrier in the $i$-th time slot, ${{\bf{F}}_{{\rm{RF}},i}} = \left[ {{{\bf{f}}_{i,1}},...,{{\bf{f}}_{i,{N_{{\rm{RF}}}}}}} \right] \in \mathbb{C}^{M \times N_{\rm{RF}}}$ is the RF part of the pilot signal with its colums denoted by ${{\bf{f}}_{i,n}} \in \mathbb{C}^{M \times 1}$, $n = 1,...,N_{\rm{RF}}$, and ${{\bf{f}}_{{\rm{BB}},i,k}} \in \mathbb{C}^{N_{\rm{RF}} \times 1}$ is the BB part of the pilot signal. Here we assume that ${{\bf{F}}_{{\rm{RF}},i}}$ is independent to different subcarriers \cite{GaoCL}. The entries of ${{\bf{F}}_{{\rm{RF}},i}}$ are of constant modulus that ${\left| {{{\left[ {{{\bf{F}}_{{\rm{RF}},i}}} \right]}_{m,n}}} \right|^2} = {M^{ - 1}}$, $\forall m \in \{1,...,M\}$, $\forall n \in \{1,...,N_{\rm{RF}}\}$.
At the user, the received pilot signal at the $k$-th subcarrier from the $i$-th time slot can be expressed as
\begin{align}
{y_{i,k}} & = {\bf{h}}_{{\rm{d}},k}^T{{\bf{s}}_{i,k}} + {\bf{h}}_{{\rm{r}},k}^T{{\bf{\Theta }}_i}{{\bf{G}}_{k}}{{\bf{s}}_{i,k}} + n_{i,k} \nonumber \\
& = [{\bf{h}}_{{\rm{d}},k}^T,{\bf{h}}_{{\rm{r}},k}^T]\left[ {\begin{array}{*{20}{c}}
{{{\bf{s}}_{i,k}}}\\
{{{\bf{\Theta }}_i}{{\bf{G}}_{k}}{{\bf{s}}_{i,k}}}
\end{array}} \right] + n_{i,k} \text{,}
\end{align}
where $n_{i,k} \sim {\cal C}{\cal N}(0,\sigma_{\rm n}^2)$ is the additive white Gaussian noise (AWGN) and ${\bf{\Theta }}_i = {{\rm diag}}\left( {{{\bm{\theta}}_i}} \right) \in \mathbb{C}^{N \times N}$ is a diagonal matrix whose digonal elements ${{\bm{\theta}}_i} \in \mathbb{C}^{N \times 1}$ represent the phase compensations brought by IRS to the signals. The elements in ${{\bm{\theta}}_i}$ are restrained by $\left| {{{\left[ {{\bm{\theta} _i}} \right]}_n}} \right| = 1$, $\forall n \in \{1,...,N\}$.

Collecting $\{ {y_{i,k}}\} _{i = 1}^{{N_{\rm{P}}}}$ in $N_{\rm P}$ successive OFDM symbols, we have
\begin{align}
\label{equ:CSpro1}
{\bf{y}}_k^T & = [{\bf{h}}_{{\rm{d}},k}^T,{\bf{h}}_{{\rm{r}},k}^T]\left[ {\begin{array}{*{20}{c}}
{{{\bf{s}}_{1,k}}}\\
{{{\bf{\Theta }}_1}{{\bf{G}}_{k}}{{\bf{s}}_{1,k}}}
\end{array}\begin{array}{*{20}{c}}
 \cdots &{{{\bf{s}}_{{N_{\rm{p}}},k}}}\\
 \cdots &{{{\bf{\Theta }}_{{N_{\rm{p}}}}}{{\bf{G}}_{k}}{{\bf{s}}_{{N_{\rm{p}}},k}}}
\end{array}} \right] \nonumber \\
& \quad + [{n_{1,k}},...,{n_{{N_{\rm{p}}},k}}] \nonumber \\
& = [{\bf{h}}_{{\rm{d}},k}^T,{\bf{h}}_{{\rm{r}},k}^T]{\bf{\Phi }}_k^T + {\bf{n}}_k^T \text{,}
\end{align}
where ${\bf{y}}_k = [{y_{1,k}},...,{y_{{N_{\rm{p}}},k}}]^T$ , ${\bf{n}}_k = [{n_{1,k}},...,{n_{{N_{\rm{p}}},k}}]^T$, and
\begin{equation}
{\bf{\Phi }}_k =
\left[ {\begin{array}{*{20}{c}}
{{{\bf{s}}_{1,k}}}\\
{{{\bf{\Theta }}_1}{{\bf{G}}_{k}}{{\bf{s}}_{1,k}}}
\end{array}\begin{array}{*{20}{c}}
 \cdots &{{{\bf{s}}_{{N_{\rm{p}}},k}}}\\
 \cdots &{{{\bf{\Theta }}_{{N_{\rm{p}}}}}{{\bf{G}}_{k}}{{\bf{s}}_{{N_{\rm{p}}},k}}}
\end{array}} \right]^T \nonumber
\end{equation}
are the aggregate received pilot signal, the aggregate AWGN vector and the measurement matrix, respectively. We rewrite (\ref{equ:CSpro1}) as a canonical form of estimation problem that
\begin{equation}
\label{equ:CSpro2}
{{\bf{y}}_k} = {{\bf{\Phi }}_{k}}{{\bf{h}}_{{\rm{eff}},k}} + {{{{\bf{n}}}_k}} \text{,}
\end{equation}
where ${{\bf{h}}_{{\rm{eff}},k}} = {[{\bf{h}}_{{\rm{d}},k}^T,{\bf{h}}_{{\rm{r}},k}^T]^T} \in \mathbb{C}^{(M+N) \times 1}$ is the effective channel vector to be estimated later.

\subsection{Signal Model with the Prior Knowledge of IRS}
We present the signal model with the knowledge of the BS-to-IRS channel ${\bf G}_k$. In practical mmWave communication systems, the assumption of the prior knowledge of ${\bf G}_k$ is reasonable, which can be explained as follows.

{\it Remark 1: There always exists a LoS path between the BS and IRS. BS can leverage the position of IRS to construct the LoS part of the BS-to-IRS channel for CE.}

As the IRS is usually integrated into the walls of the building \cite{Fellow3} without mobility and the BS is elevated high with few obstacles around \cite{GaoTSP}, a stable LoS-path transsmision between the BS and IRS is expected. The prior knowledge of the positioning of the IRS can help to easily obtain the directions of LoS transmission from the BS to IRS (i.e., AoA $\{\theta _{{\rm{g}},0}^{\rm{r}},\varphi _{{\rm{g}},0}^{\rm{r}}\}$ and AoD $\{\theta _{{\rm{g}},0}^{\rm{t}},\varphi _{{\rm{g}},0}^{\rm{t}}\}$). Those parameters can be delivered to users via control channel to construct the LoS part of the BS-to-IRS channel\footnote{Note that the parameter ${{\beta _{{\rm{g}},0,k}}}$ in ${\bf{G}}_{{\rm{L}},k}$ can be included into the effective channel vector ${{\bf{h}}_{{\rm{eff}},k}}$ to be estimated. For simplicity, we assume the knowledge of the complete ${\bf{G}}_{{\rm{L}},k}$ at the users without loss of generality.}.

{\it Remark 2: The path loss for NLoS paths is much larger than that for LoS path in mmWave systems. Thus, the LoS path will dominate the energy of the channel and the NLoS part can be neglected.}

In mmWave systems, the typical value of Rician factor of the channels, $K_{\rm f}$, is $20$ dB \cite{GaoCL} and can be up to $40$ dB in some cases \cite{Std}, which is sufficiently large to validate {\it Remake 2}.

Based on the analysis above, we can substitute (\ref{equ:Gf2}) into (\ref{equ:CSpro1}) and treat the items associated with ${{{\bf{G}}_{{\rm{N}},k}}}$ as additive measurement noise, that is
\begin{align}
\label{equ:CSpro3}
&{\bf{y}}_k^T  = [{\bf{h}}_{{\rm{df}},k}^T,{\bf{h}}_{{\rm{rf}},k}^T]\underbrace {\left[ {\begin{array}{*{20}{c}}
{{{\bf{s}}_{1,k}}}\\
{{{\bf{\Theta }}_1}{{\bf{G}}_{{\rm{L}},k}}{{\bf{s}}_{1,k}}}
\end{array}\begin{array}{*{20}{c}}
 \cdots &{{{\bf{s}}_{{N_{\rm{p}}},k}}}\\
 \cdots &{{{\bf{\Theta }}_{{N_{\rm{p}}}}}{{\bf{G}}_{{\rm{L}},k}}{{\bf{s}}_{{N_{\rm{p}}},k}}}
\end{array}} \right]}_{{{\bf{\Phi }}_{{\rm{L}},k}}} \nonumber \\
& + \underbrace {{\bf{h}}_{{\rm{rf}},k}^T\left[ {{{\bf{\Theta }}_1}{{\bf{G}}_{{\rm{N}},k}}{{\bf{s}}_{1,k}},...,{{\bf{\Theta }}_{{N_{\rm{p}}}}}{{\bf{G}}_{{\rm{N}},k}}{{\bf{s}}_{1,k}}} \right] + [{n_{1,k}},...,{n_{{N_{\rm{p}}},k}}]}_{{{{\bf{\bar n}}}_k}} \text{.}
\end{align}
Note that the measurement matrix under the prior knowledge of IRS ${\bf{\Phi }}_{\rm{L},k} \in \mathbb{C}^{N_{\rm p} \times (M+N)}$, $\forall k \in \{1,...,K\}$ are known to the users for CE due to the prior knowledge of the LoS part of the BS-to-IRS channel. (\ref{equ:CSpro3}) can be rewritten as
\begin{equation}
\label{equ:CSLoS}
{{\bf{y}}_k} = {{\bf{\Phi}}_{{\rm{L}},k}}{{\bf{h}}_{{\rm{eff}},k}} + {{{{\bf{\bar n}}}_k}} \text{,}
\end{equation}
where ${{{{\bf{\bar n}}}_k}}$ is the effective measurement noise vector.

For mmWave massive MIMO systems, we usually have $(M+N) > N_{\rm p}$ because of large dimension of antenna arrays and the limited channel coherence time, which makes (\ref{equ:CSLoS}) an under-determined system and brings great challenge to conventional CE techniques, such as LS estimator or linear minimum mean square error (LMMSE) estimator. Fortunately, by leveraging the well-known angular sparsity of mmWave channels \cite{GaoCL, BShim,Mag}, one can efficiently solve the under-determined problem (\ref{equ:CSLoS}) under the framework of CS technique, which will be clarified later in this paper.

\section{Compressive Sensing Based Channel Estimation Algorithm}
Facing the difficulty in solving under-determined system (\ref{equ:CSLoS}), extra information is inevitable to be exploited. Fortunately, the state-of-the-art CS techniques can be introduced to conduct CE in IRS-aided mmWave massive MIMO systems.

\subsection{Pilot Design for IRS-Aided Systems}
Careful design of the pilot signals is essential to guarantee the performance of recovering the channels \cite{GaoCL, GaoTSP, BShim}. Considering the unique signal model in (\ref{equ:CSpro1}) brought by IRS, the pilot signals is essentially composed of two parts: the pilot signal ${{{\bf{s}}_{i,k}}}$ at the BS and the phase compensations ${{{\bf{\Theta }}_i}}$ at the IRS. In this paper, we focus on the design of the RF part of ${{{\bf{s}}_{i,k}}}$ and set ${{\bf{f}}_{{\rm{BB}},i,k}} = \sqrt {\frac{{{P_{{\rm{Tx}}}}}}{{{N_{{\rm{RF}}}}}}} {\bf{1}} \in \mathbb{C} ^{{N_{{\rm{RF}}}} \times 1}$ for an indiscriminate power allocation among all RF chains, where $P_{{\rm Tx}}$ is the total transmit power. Thus, the pilot signal at the BS in (\ref{equ:S}) turns into
\begin{equation}
\label{equ:S2}
{{\bf{s}}_{i,k}} = \sqrt{\frac{{P_{{\rm{Tx}}}}}{{ {{N_{{\rm{RF}}}}} }}}\sum\limits_{r = 1}^{{N_{{\rm{RF}}}}} {{{\bf{f}}_{i,r}}} \text{,}
\end{equation}
which consists of ${{N_{{\rm{RF}}}}}$ RF components. Given that ${{\bf{s}}_{i,k}}$ will be transmitted in both the BS-to-user channel and BS-to-IRS channel, the design of ${{\bf{s}}_{i,k}}$ should take into account the characteristics of both the channels. We first decompose all the RF components in (\ref{equ:S2}) into two parts as
%\begin{equation}
%\label{equ:S3}
${{\bf{s}}_{i,k}} = {\bf{s}}_{i,k}^{{\rm{U}}} + {\bf{s}}_{i,k}^{{\rm{I}}}$,
%\end{equation}
where
\begin{equation}
{\bf{s}}_{i,k}^{{\rm{U}}} = \sqrt{\frac{{P_{{\rm{Tx}}}}}{{ {{N_{{\rm{RF}}}}} }}}\sum\limits_{r^{\rm U} = 1}^{{N_{{\rm{RF}}}^{{\rm U}}}} {{{\bf{f}}_{i,r^{\rm U}}}} \ \ \text{and} \ \ {\bf{s}}_{i,k}^{{\rm{I}}} = \sqrt{\frac{{P_{{\rm{Tx}}}}}{{ {{N_{{\rm{RF}}}}} }}}\sum\limits_{r^{\rm I} = 1}^{{N_{{\rm{RF}}}^{{\rm I}}}} {{{\bf{f}}_{i,{N_{{\rm{RF}}}^{{\rm U}}}+r^{\rm I}}}} \nonumber
\end{equation}
are the pilot signals dedicated to the user and IRS, respectively, ${N_{{\rm{RF}}}^{{\rm U}}}$ and ${N_{{\rm{RF}}}^{{\rm I}}}$ are the numbers of RF components allocated to ${\bf{s}}_{i,k}^{{\rm{U}}}$ and ${\bf{s}}_{i,k}^{{\rm{I}}}$, respecively, with ${N_{{\rm{RF}}}} = {N_{{\rm{RF}}}^{{\rm U}}} + {N_{{\rm{RF}}}^{{\rm I}}}$.

{\it 1) Design of ${\bf{s}}_{i,k}^{{\rm{I}}}$ Based on the Prior Knowlegde of IRS}: Though the IRS can provide the virtual LoS transmission between the BS and blocked user, the signals reflected by the IRS will go through much longer transmission distance. Thus, extremely high path loss will severely degrade the power of the received pilot signals. Fortunately, the beamforming gain brought by the massive MIMO at the BS can compensate this high path loss when the optimal beam pair between the BS and IRS is found and aligned. Considering that the positioning of IRS is pre-known to the BS as stressed in Section III-B, the optimal beam direction at the BS is exactly the known AoD $\{\theta _{{\rm{g}},0}^{\rm{t}},\varphi _{{\rm{g}},0}^{\rm{t}}\}$. Accordingly, ${\bf{s}}_{i,k}^{{\rm{I}}}$, $\forall i$ are designed as the following closed-form expression:
\begin{equation}
\label{equ:SI}
{\bf{s}}_{i,k}^{{\rm{I}}} = \sqrt {\frac{{{P_{{\rm{Tx}}}}}}{{{N_{{\rm{RF}}}}}}} N_{{\rm{RF}}}^{\rm{I}} \cdot  {\bf{a}}_M(\theta _{{\rm{g}},0}^{\rm{t}},\varphi _{{\rm{g}},0}^{\rm{t}}) \text{,} \  \forall i \text{,}
\end{equation}
with each ${{{\bf{f}}_{i,r^{\rm I}}}}$, $r^{\rm I} = 1,...,N_{{\rm{RF}}}^{\rm{I}}$ assigned as ${\bf{a}}_M(\theta _{{\rm{g}},0}^{\rm{t}},\varphi _{{\rm{g}},0}^{\rm{t}})$. The designed pilot signals in (\ref{equ:SI}) will provide the beamforming gain proportional to the number of antennas at the BS ($M$), which helps to significantly enhance the power of the received pilot signals and estimate the IRS-to-uesr channel ${\bf{h}}_{{\rm{r}},k}^T$ more accurately;

{\it 2) Design of ${\bf{s}}_{i,k}^{{\rm{U}}}$ and ${\bf{\Theta }}_i = {{\rm diag}}\left( {{{\bm{\theta}}_i}} \right)$}: Since no prior information of the user can be exploited at the BS as well as IRS, the related pilot signals ${\bf{s}}_{i,k}^{{\rm{U}}}$ and ${\bf{\Theta }}_i$ should be as diversified as possible to guarantee the complete sounding of the channels. Specifically, the ${\bf{s}}_{i,k}^{{\rm{U}}}$ and ${\bf{\Theta }}_i = {{\rm diag}}\left( {{{\bm{\theta}}_i}} \right)$ are designed as
\begin{equation}
\label{equ:SU}
{\left[ {{{\bf{f}}_{i,N_{{\rm{RF}}}^{\rm{U}} + {r^{\rm{I}}}}}} \right]_m} = \frac{1}{M}{e^{j{\phi _{i,{r^{\rm{I}}},m}}}} \text{,}
\end{equation}
\begin{equation}
\label{equ:theta}
\left| {{{\left[ {{\bm \theta _i}} \right]}_n}} \right| = {e^{j{\phi _{i,n}}}} \text{,}
\end{equation}
where the random variables ${{\phi _{i,{r^{\rm{I}}},m}}}$ and ${{\phi _{i,n}}}$ follow the independent identically distributed uniform distribution ${\cal U}\left[ {0,2\pi } \right)$.

\subsection{Sparse Representation and Distributed OMP Algorithm}
We note that the mmWave transmission with extremely high carrier frequency suffers from severely high path loss and blockage effect. Consequently, there will be only a few multipath components (with different AoAs/AoDs) between the transmitter and receiver, resulting in the well-known {\it angular sparsity of the mmWave channels} \cite{GaoCL, BShim}. We consider the virtual angular domain representation \cite{GaoTSP} to elaborate this sparsity of the channels, based on which ${\bf{h}}_{{\rm{d}},k}^T$ can be rewritten as
\begin{equation}
\label{equ:hda}
{\bf{h}}_{{\rm{d}},k}^T = {\bf{h}}_{{\rm{da}},k}^T{\bf{A}}_{\rm{d}}^H + {{\bf{e}}^T_{{\rm{d}},k}} \text{,}
\end{equation}
where ${\bf{A}}_{\rm{d}}$ is the transformation matrix consisting of some {\it samples} of virtual angles, ${\bf{h}}_{{\rm{da}},k}$ is the representation of ${\bf{h}}_{{\rm{d}},k}$ in the virtual angular domain, and ${{\bf{e}}_{{\rm{d}},k}}$ is the quantization error vector caused by the mismatch between the real angles and the samples in ${\bf{A}}_{\rm{d}}$. Given the angular sparsity of the mmWave channels, ${\bf{h}}_{{\rm{da}},k}$ will exhibit the sparsity. Similarly, ${\bf{h}}_{{\rm{r}},k}^T$ can be expressed as
\begin{equation}
\label{equ:hra}
{\bf{h}}_{{\rm{r}},k}^T = {\bf{h}}_{{\rm{ra}},k}^T{\bf{A}}_{\rm{r}}^H + {{\bf{e}}^T_{{\rm{r}},k}} \text{.}
\end{equation}

\begin{algorithm}[t]
\caption{Distributed OMP algorithm}
\label{alg:alg1}
\begin{algorithmic}[1]
\renewcommand{\algorithmicrequire}{\textbf{Input:}}
\renewcommand{\algorithmicensure}{\textbf{Output:}}
\Require {The noise-polluted measurements ${{\bf{y}}_k}$, the sensing matrices ${{\bf{\Phi }}_{{\rm{L},}k}}{\bf{\Psi}}$, and the threshold $\varepsilon$ for the stop criterion.}
\Ensure {The estimated channels ${\bf{\hat h}}_{{\rm{d}},k}$ and ${\bf{\hat h}}_{{\rm{r}},k}$.}
\State Initialization: $\mathcal{I} = $ empty set, ${{\bf{r}}_k} = {{\bf{y}}_k}$ and ${\bf{\hat h}}_{{\rm{effa}},k} = {\bf{0}}$.
\While{$\frac{1}{{K{N_{\rm{P}}}}}\sum\limits_{k = 1}^K {\left\| {{{\bf{r}}_k}} \right\|_2^2} > \varepsilon$,}
    \State ${i^*} = \mathop {\arg \max } \limits_i \sum\limits_{k = 1}^K \left| {{{\left[ { {{{\left( {{{\bf{\Phi }}_{{\rm{L}},k}}{\bf{\Psi }}} \right)}^H}{{\bf{r}}_k}} } \right]}_i}} \right|$;
    \State ${{\mathcal{I}}}={{\mathcal{I}}} \cup \{{{i}^{*}}\}$;
    \State ${\bf{\hat h}}_{k} = \left[ {{{\bf{\Phi }}_{{\rm{L}},k}}{\bf{\Psi }}} \right]_{\cal I}^\dag {{\bf{y}}_k}$;
    \State ${{\bf{r}}_k} = {{\bf{y}}_k} - \left[ {{\bf{\Phi }}_{{\rm{L}},k}}{\bf{\Psi }}\right]_{\cal I}{{\bf{\hat h}}_k}$;
\EndWhile
\vspace{1mm}
\State ${\left[ {{{{\bf{\hat h}}}_{{\rm{effa}},k}}} \right]_{\cal I}} = {{{\bf{\hat h}}}_k}$;
\State Decompose ${{{{\bf{\hat h}}}_{{\rm{effa}},k}}}$ as ${{{{\bf{\hat h}}}_{{\rm{effa}},k}}} = [{\bf{\hat h}}_{{\rm{da}},k}^T,{\bf{\hat h}}_{{\rm{ra}},k}^T]^T$;
\State ${\bf{\hat h}}_{{\rm{d}},k}^T = {\bf{\hat h}}_{{\rm{da}},k}^T{\bf{A}}_{\rm{d}}^H$, ${\bf{\hat h}}_{{\rm{r}},k}^T = {\bf{\hat h}}_{{\rm{ra}},k}^T{\bf{A}}_{\rm{r}}^H$
\end{algorithmic}
\end{algorithm}

We substitute (\ref{equ:hda}) and (\ref{equ:hra}) into (\ref{equ:CSLoS}) to obtain
\begin{align}
\label{equ:CSlast}
{{\bf{y}}_k} & = {{\bf{\Phi }}_{{\rm{L},}k}} \left( {[{\bf{h}}_{{\rm{da}},k}^T{\bf{A}}_{{\rm d}}^H,{\bf{h}}_{{\rm{ra}},k}^T{\bf{A}}_{{\rm r}}^H]^T} + {[{\bf{e}}_{{\rm{d}},k}^T,{\bf{e}}_{{\rm{r}},k}^T]^T} \right) + {{{{\bf{\bar n}}}_k}} \nonumber \\
& = {{\bf{\Phi }}_{{\rm{L},}k}}{\underbrace {\left[ {\begin{array}{*{20}{c}}
{{\bf{A}}_{{\rm d}}^H}&{}\\
{}&{{\bf{A}}_{{\rm r}}^H}
\end{array}} \right]}_{\bf{\Psi }}}^T {\underbrace{[{\bf{h}}_{{\rm{da}},k}^T,{\bf{h}}_{{\rm{ra}},k}^T]^T}_{{{\bf{h}}_{{\rm{effa}},k}}}} + {{\bf{n}}_{{\rm{eff}},k}} \nonumber \\
& = {{\bf{\Phi }}_{{\rm{L},}k}}{\bf{\Psi}}{{\bf{h}}_{{\rm{effa}},k}} + {{\bf{n}}_{{\rm{eff}},k}} \text{,}
\end{align}
where ${{\bf{h}}_{{\rm{effa}},k}}$ is the effective sparse channel vector to be estimated, ${\bf{\Psi}} $ is called effective dictionary matrix which simultaneously sparsifies ${\bf{h}}_{{\rm{d}},k}$ and ${\bf{h}}_{{\rm{r}},k}$, and ${{\bf{n}}_{{\rm{eff}},k}} = {{\bf{\Phi }}_{{\rm{L},}k}}{[{\bf{e}}_{{\rm{d}},k}^T,{\bf{e}}_{{\rm{r}},k}^T]^T} + {{{{\bf{\bar n}}}_k}}$ is the effective noise vector taking both the measurement noise and the quantization error into account.

Since the spatial propagation characteristics of the channels within the system bandwidth are almost unchanged, the subchannels at different subcarriers will share the common sparsity in the virtual angular domain \cite{GaoCL, GaoTSP}, namely,
\begin{equation}
\label{equ:supp}
{\rm{supp(}}{{\bf{h}}_{{\rm{effa}},1}}{\rm{)}} = ... = {\rm{supp(}}{{\bf{h}}_{{\rm{effa}},K}}{\rm{)}} \text{.}
\end{equation}
The task of CE turns into solving (\ref{equ:CSlast}) for the sparse solution of ${{\bf{h}}_{{\rm{effa}},k}}$ under the constraint in (\ref{equ:supp}). Then ${\bf{h}}_{{\rm{d}},k}$ and ${\bf{h}}_{{\rm{r}},k}$ can be reconstructed based on the estimated ${{{\bf{h}}_{{\rm{effa}},k}}}$ according to (\ref{equ:hda}) and (\ref{equ:hra}) (ignore the quantization errors), respectively.

An efficient CS-based algorithm called distributed orthogonal matching pursuit (DOMP) to solve (\ref{equ:CSlast}) is provided in {\bf Algorithm 1}. Compared with the simple orthogonal matching pursuit \cite{BShim}, the main improvement of DOMP lies in the line 3 in {\bf Algorithm 1}, where all $K$ subchannels are jointly exploited to detect the support of the channels in the virtual angular domain.

% 左右并排仿真图（方法一）
\begin{figure}[b]
\vspace*{-4mm}
	\label{fig:compare}
     \centering
     \hspace{-0.1in}
     \subfigure[]
     {
     	\label{fig:leakage}
     	\centering
	    \includegraphics[width=1.65in, keepaspectratio]%
	    {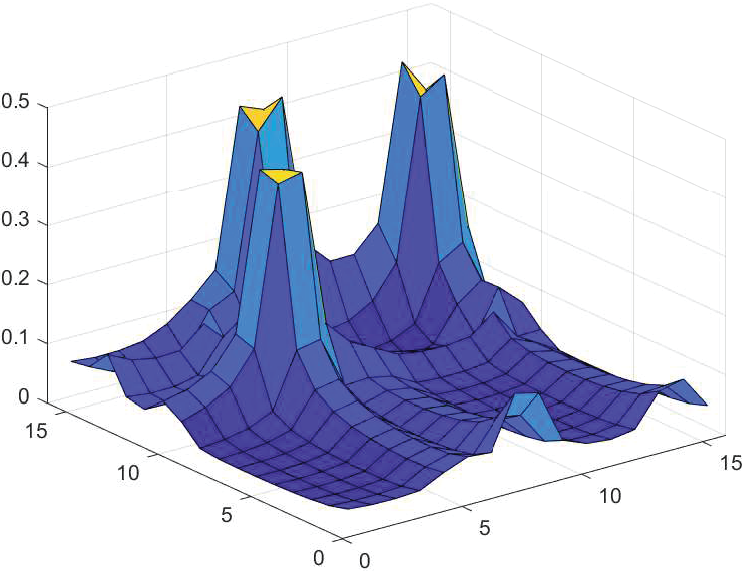}
	    \vspace*{-1.5mm}
     }
     \hspace{-0.05in}
     \subfigure[]
     {
     	\label{fig:pure}
     	\centering
	    \includegraphics[width=1.65in, keepaspectratio]%
	    {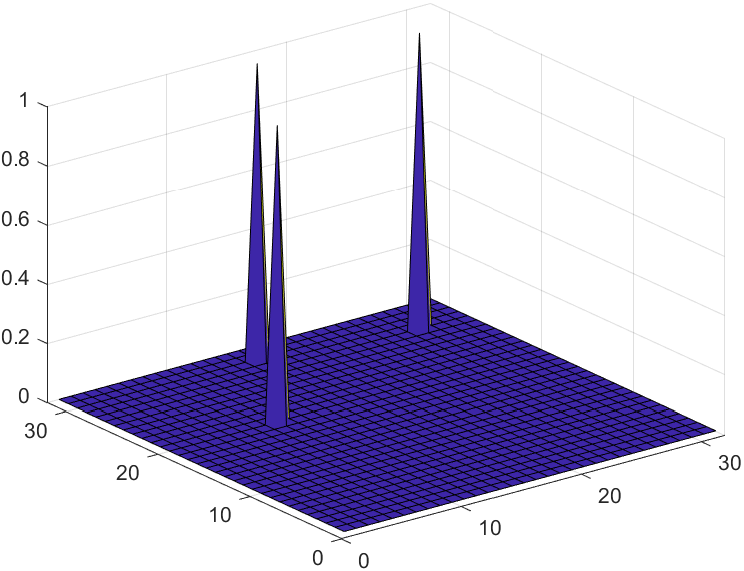}
	    \vspace*{-1.5mm}
     }
     \caption{The sparse representations of the channel in the virtual angular domain. (a) unitary scenario with degraded sparsity caused by power leakage; (b) redundant scenario with enhanced sparsity.}
%     \vspace*{-4mm}
\end{figure}

\subsection{Redundant Dictionary Design}
Note that the transformation matrices ${\bf{A}}_{{\rm{d}}}$ in (\ref{equ:hda}) and ${\bf{A}}_{{\rm{r}}}$ in (\ref{equ:hra}) (also called {\it dictionaries}) are determined by the geometrical structure of the antenna arrays \cite{GaoTSP}. In some previous work \cite{GaoTSP, GaoCL}, the transformation matrix is designed under the constraint of unitarity to guarantee the uniqueness of the transformation between the spatial domain and virtual angular domain. However, this kind of transformation matrix suffers from the limited resolution in the virtual angular domain. The angles of the real channel will mismatch the samples in the virtual angular domain with high probability. This behavior causes the {\it power leakage} and degrades the sparsity of ${\bf{h}}_{{\rm{da}},k}$ (see Fig. \ref{fig:leakage}), and thus makes the quantization error ${{\bf{e}}_{{\rm{d}},k}}$ non-negligible. For the improvement, we ignore the unitarity of the transformation matrix and design it as
\begin{equation}
\label{equ:redu}
{\bf{A}}_{\rm{d}} = {{\bf{A}}_{M,G_M}} = \sqrt {\frac{M}{{{G_M}}}} ({{\bf{F}}_{{G_{M_x}}}}{)_{1:{M_x}}} \otimes {{\rm{(}}{{\bf{F}}_{{G_{M_y}}}})_{1:{M_y}}} \text{,}
\end{equation}
where the notation ${({{\bf{F}}_G})_{1:M}}$ represents the submatrix of ${{\bf{F}}_G}$ constructed from the first $M$ rows of ${{\bf{F}}_G}$, $G_{M_x}$ and $G_{M_y}$ are the numbers of equally-spaced samples along the horizontal and vertical direction, respectively, and $G_M = G_{M_x} G_{M_y}$.
Instuitively, with $G_{M_x} > M_x$ or $G_{M_y} > M_y$, the angles of the real channel are more likely to coincide with some samples in ${{\bf{A}}_{M,G_M}}$. Therefore, the quantization error ${{\bf{e}}_{{\rm{d}},k}}$ can be further suppressed and treated as ignorable minor noise. We refer to ${\bf{A}}_{M,G_M}$ as the {\it redundant dictionary} \cite{My, ALiao} when $G_M > M$. For the same reason, we can design ${\bf{A}}_{\rm{r}}$ as ${{\bf{A}}_{N,G_N}}$ with $G_N = G_{N_x} G_{N_y} > N$.

In Fig. 2, we provide an example of the sparse representations of ${\bf{h}}_{{\rm{d}},k}$ with $3$ paths in the virtual angular domain under the unitary ($M_x = M_y = 16$) and redundant ($G_{M_x} = G_{M_y} = 32$) scenario. It is clear that the redundant dictionary can enhance the sparsity of the channel, thus the improved performance of the CS-based CE is expected, which will be demonstrated in the simulation following.

\section{Simulation Results}
We consider the standard urban micro (UMi)-street canyon scenario described by 3GPP in \cite{Std}. In the experiments, the system is with carrier frequency $30$ GHz and bandwidth $100$ MHz, $M = M_x \times M_y = 16 \times 16$, $N = N_x \times N_y = 16 \times 16$, $K = N_{\rm CP} = 64$, and the number of channel NLoS paths is $6$. The Rician factor is set as $K_{\rm f} = 20$ dB. The raised cosine filter with roll-off factor $0.8$ is adopted as $p(\tau)$. We set $N_{\rm RF} = 2$ and $N_{\rm RF}^{\rm I} = N_{\rm RF}^{\rm U} = 1$.
The noise-power spectrum density at the user is $\sigma _{{\rm{NSD}}}^2 = -174$ dBm/Hz, and the noise power for each subchannel can be accordingly calculated. We empirically set $\varepsilon = \sigma _{\rm{n}}^2$ in {\bf Algorithm 1}.
Since the blocked and unblocked users (see Fig. \ref{fig:model}) have different characteristics of the channels, we will investigate them respectively in the sequel. We assume that the blocked (unblocked) users communicate with BS (IRS) by only the NLoS paths.
We define ${r_{\rm{p}}} \buildrel \Delta \over = \frac{{{N_{\rm{P}}}}}{{M + N}}$ to evaluate the pilot overhead of the proposed CS-based CE scheme, and ${r_{{\rm{dic}}}} \buildrel \Delta \over = \frac{{{G_M}}}{M} = \frac{{{G_N}}}{N}$ to evaluate the redundant dictionary design.

In Fig. 3, we plot the normalized mean square error (NMSE) performances of estimating ${\bf{ h}}_{{\rm{r}},k}$ as a function of transmit power $P_{\rm Tx}$ for the blocked users\footnote{For a blocked user, the signals from the NLoS channel ${\bf{ h}}_{{\rm{d}},k}$ can be treated as noise. Therefore, we only preserve the NMSE performances of estimating ${\bf{h}}_{{\rm{r}},k}$ in Fig. 3. For the same reason, we only preserve the NMSE performances of estimating ${\bf{h}}_{{\rm{d}},k}$ in Fig. 4 for the unblocked users.}. For benchmark, we weigh in the conventional well-determined LS estimator \cite{LSCE} without the receive noise, i.e., solving (\ref{equ:CSLoS}) via LS algorithm when $N_p = M+N$ ($r_{\rm p}=1$). It can be seen from Fig. 3(a) that even with much higher pilot overhead and the absence of the receive noise, the performance of LS estimator is limited by the interference brought by the NLoS paths of the BS-to-IRS channel. However, the proposed CS-based CE scheme can leverage the sparsity of the channels and thus has better robustness against the additive noise or interference. The NMSE performance of the proposed CE scheme can exceed that of the LS estimator when $P_{\rm Tx}$ increases. The similar results can be observed in Fig. 4(a) for the unblocked users, which further verifies the superiority of the proposed scheme over the conventional CE technique.

% 左右并排仿真图（方法一）
\begin{figure}[t]
     \centering
     \hspace{-0.1in}
     \subfigure[]
     {
%     	\label{fig:temp1}
     	\centering
	    \includegraphics[width=1.65in, keepaspectratio]%
	    {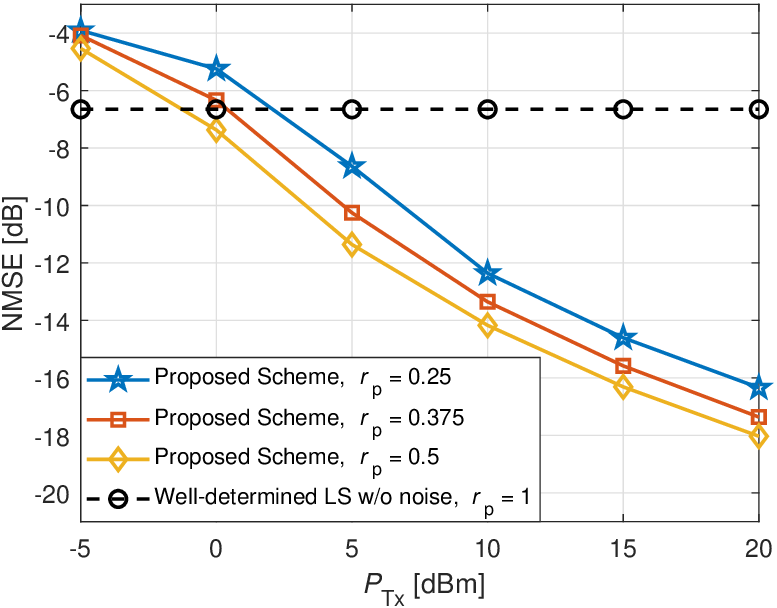}
	    \vspace*{-1.5mm}
     }
     \subfigure[]
     {
%     	\label{fig:temp2}
     	\centering
	    \includegraphics[width=1.65in, keepaspectratio]%
	    {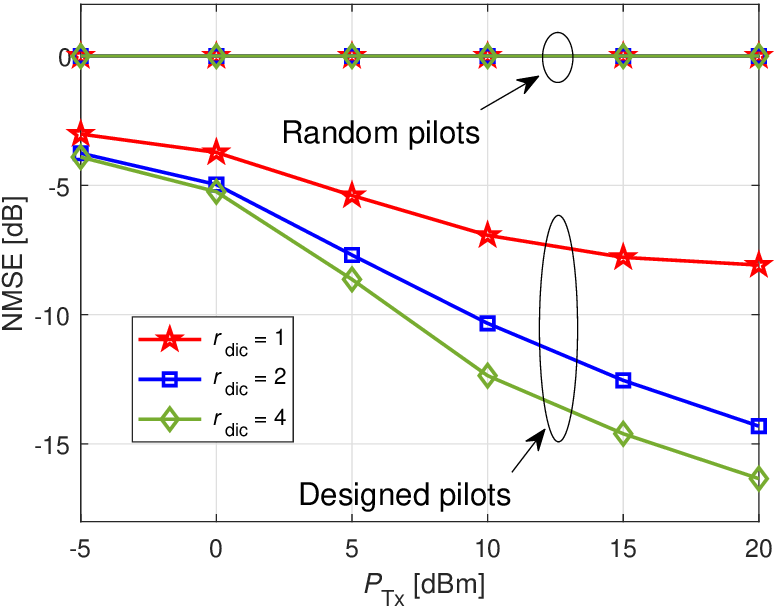}
	    \vspace*{-1.5mm}
     }
     \caption{NMSE performances of estimating ${\bf{ h}}_{{\rm{r}},k}$ for the blocked users. (a) $r_{\rm dic}$ is fixed as $4$; (b) $r_{\rm p}$ is fixed as $0.25$.}
     \vspace*{-2mm}
\end{figure}

% 左右并排仿真图（方法一）
\begin{figure}[t]
     \centering
     \hspace{-0.1in}
     \subfigure[]
     {
%     	\label{fig:temp1}
     	\centering
	    \includegraphics[width=1.65in, keepaspectratio]%
	    {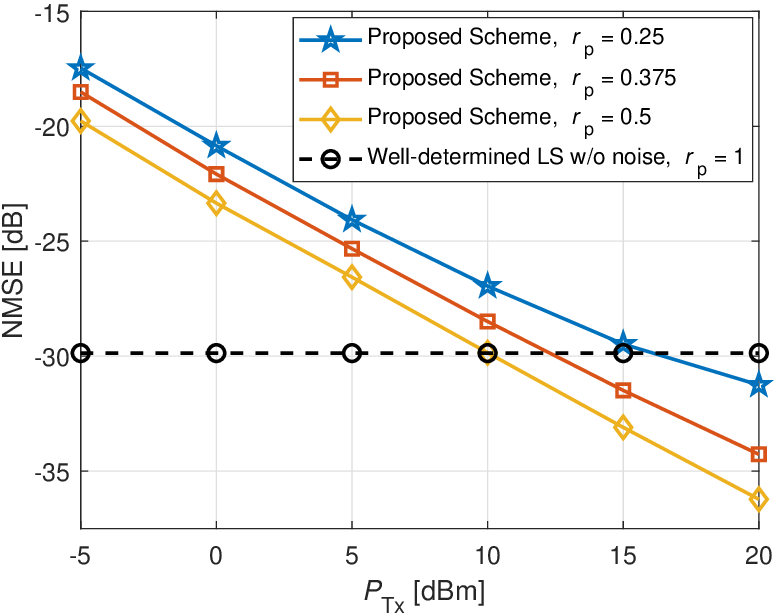}
	    \vspace*{-1.5mm}
     }
     \subfigure[]
     {
%     	\label{fig:temp2}
     	\centering
	    \includegraphics[width=1.65in, keepaspectratio]%
	    {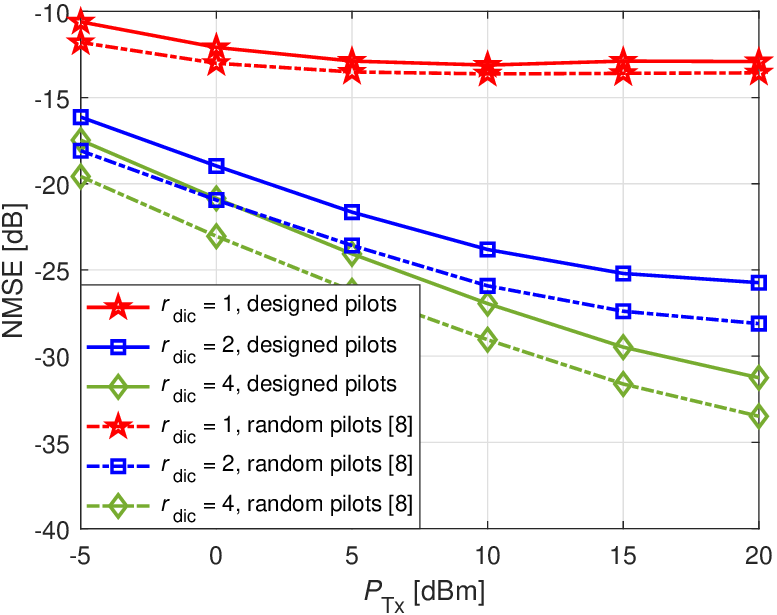}
	    \vspace*{-1.5mm}
     }
     \caption{NMSE performances of estimating ${\bf{ h}}_{{\rm{d}},k}$ for the unblocked users. (a) $r_{\rm dic}$ is fixed as $4$; (b) $r_{\rm p}$ is fixed as $0.25$.}
     \vspace*{-2mm}
\end{figure}

To investigate the effectiveness of the proposed pilot design and the redundant dictionary design, the NMSE performances of the proposed scheme under different settings are provided in Fig. 3(b) and Fig. 4(b) for the blocked and unblcoked users, respectively. It can be seen that the redundant dictionary ($r_{\rm dic} > 1$) can significantly improve CE performance compared with the unitary dictionary ($r_{\rm dic} = 1$). Interesting results are observed when investigating the effectiveness of the proposed pilot design. We consider the fully random pilot design in \cite{GaoCL} for comparison. On the one hand, for the blocked users, the signals reflected by IRS suffers from severe path loss. Thus, the proposed pilot design which provides high beamforming gain can outperform the fully random pilot design in \cite{GaoCL} substantially, as illustrated in Fig. 3(b).
On the other hand, we observe from Fig. 4(b) that the proposed pilot design has a minor negative effect on estimating ${\bf{ h}}_{{\rm{d}},k}$ for the unblocked users. It is reasonable because the proposed pilot design allocates part of the RF resources for IRS and thus it decreases the power of pilot signals transmitted in ${\bf{ h}}_{{\rm{d}},k}$. Considering that the status (blocked or unblocked) of the users cannot be known before CE, the proposed CS-based CE solution is still an appealing techique for IRS-aided mmWave massive MIMO systems.

\section{Conclusions}
We have proposed a CS-based broadband CE solution for IRS-aided mmWave massive MIMO systems. Specifically, we first formulated the CE for IRS-aided cellular system based on the proposed pilot training framework and the prior knowledge of IRS. Then, we proposed to decompose the pilot signals at the BS into two parts and design them respectively to guarantee the effective sounding of both the BS-to-user and IRS-to-user channels. By leveraging the angular sparsity and the common sparsity shared by all subchannels, we presented an efficient DOMP algorithm to solve the CE problem. We also designed the redundant dictionary to combat the power leakage caused by the off-grid AoAs/AoDs. From the simulation results, we demonstrated the effectiveness and superiority of the proposed CE technique.

\end{document}